\documentclass[10 pt,final,journal,letterpaper,oneside,onecolumn]{IEEEtran}
\usepackage{graphicx}
\usepackage{amssymb,amsmath,amsfonts}
\usepackage{suffix}
\usepackage{mathtools}
\usepackage{cite}  
\usepackage{amsthm}
\usepackage[utf8]{inputenc}
\usepackage[english]{babel}
\usepackage{epstopdf}
\usepackage{color}

\newtheorem{prop}{Proposition}

\DeclarePairedDelimiterX\MeijerM[3]{\lparen}{\rparen}%
{\begin{smallmatrix}#1 \\ #2\end{smallmatrix}\delimsize\vert\,#3}

\newcommand\MeijerG[8][]{%
  G^{\,#2,#3}_{#4,#5}\MeijerM[#1]{#6}{#7}{#8}}

\WithSuffix\newcommand\MeijerG*[7]{G^{\,#1,#2}_{#3,#4}\MeijerM*{#5}{#6}{#7}}                  
\ifCLASSINFOpdf
\else
\fi
\begin{document}
%
\title{Performance Analysis of Cooperative V2V and V2I Communications under Correlated Fading}


\author{Furqan Jameel, Muhammad Awais Javed, Duy T. Ngo}%

%
%

\markboth{Accepted for IEEE Transactions on Intelligent Transportation Systems}%
{Shell \MakeLowercase{\textit{et al.}}: Bare Demo of IEEEtran.cls for IEEE Journals}
%



\maketitle
\begin{abstract}
Cooperative vehicular networks will play a vital role in the coming years to implement various intelligent transportation related applications. Both vehicle-to-vehicle (V2V) and vehicle-to-infrastructure (V2I) communications will be needed to reliably disseminate information in a vehicular network. In this regard, a roadside unit (RSU) equipped with multiple antennas can improve the network capacity. While the traditional approaches assume antennas to experience independent fading, we consider a more practical uplink scenario where antennas at the RSU experience correlated fading. In particular, we evaluate the packet error probability for two renowned antenna correlation models, i.e., constant correlation (CC) and exponential correlation (EC). We also consider intermediate cooperative vehicles for reliable communication between the source vehicle and the RSU. Here, we derive closed-form expressions for packet error probability which help quantify the performance variations due to fading parameter, correlation coefficients and the number of intermediate helper vehicles. To evaluate the optimal transmit power in this network scenario, we formulate a Stackelberg game, wherein, the source vehicle is treated as a buyer and the helper vehicles are the sellers. The optimal solutions for the asking price and the transmit power are devised which maximize the utility functions of helper vehicles and the source vehicle, respectively. We verify our mathematical derivations by extensive simulations in MATLAB.
\end{abstract}


\section{Introduction}

Ubiquitous vehicular connectivity is expected to be an essential paradigm shift for guaranteeing driver safety and preventing road accidents \cite{8493115, A3}. However, with the rapid spread of information and communication technology, especially in the domain of consumer electronics, there is a need to improve different aspects of vehicular networks. Cooperative communication among vehicles is one of these aspects. Cooperative vehicular networking is a key enabler for intelligent transportation systems and smart cities. Many traffic management and passenger comfort applications can be implemented by means of efficient and reliable data exchange among vehicles \cite{8302837, A4, A5}. Although single-hop communication is typically used for the periodic exchange of mobility information among neighbor vehicles, multi-hop communications can be used to propagate emergency notifications within a large geographical area. Moreover, the multi-hop communication approach is favored when line-of-sight does not exist between source and destination; providing a mechanism to combat the attenuation of wireless signals. To ensure widespread vehicular network connectivity, a roadside unit (RSU) is placed at various strategic locations along the road \cite{7918915,jameel2019internet}. An RSU typically comprises of multiple short-range antennas \cite{charitos2017mimo} to provide uninterrupted connectivity between vehicles in the transmission range, also termed as vehicle-to-infrastructure (V2I) communications. If needed, the RSU can also act as a relay to exchange packets between two vehicles \cite{8302837,8594703}.  

While the performance limits of single-link V2I communications have been well characterized \cite{A1, A2}, only limited work has been done to investigate the performance of a multi-antenna RSU. In \cite{charalampopoulos2016v2i}, the authors considered the omnidirectional antenna at RSU to investigate the performance of V2I links in a highway scenario. By varying vehicle density, it was shown that the location of RSU was of considerable importance in vehicular communications. Moser \emph{et al}. in \cite{moser2015mimo} studied multiple antenna approaches against short-term fading vehicular communications conditions. A comparative analysis of IEEE 802.11p and IEEE 802.11p long-term evolution (LTE) HetNet was provided for multiple-input-multiple-output (MIMO) channels in \cite{charitos2017mimo, 8246040}. After the antenna radiation pattern was simulated for each scenario, it was concluded that IEEE 802.11p performs acceptably well for sparse network topologies while IEEE 802.11p LTE HetNet shows enhanced performance even in dense urban vehicular scenarios. 

Recent studies on cooperative communications have shown significant performance improvement over conventional vehicular communications. Liu \emph{et al}. in \cite{liu2016cooperative} investigated the problem of data dissemination in downlink infrastructure-to-vehicle (I2V) and vehicle-to-vehicle (V2V) communication scenarios. They analyzed the constraints and requirements of data dissemination and formulated the data scheduling problem in vehicular communications. However, they did not take into account the effect of multiple antennas at RSU and ignored direct multi-hop V2V communication among vehicles. Some studies also suggest to use energy harvesting techniques such as simultaneous wireless information and power transfer \cite{7840157,8246039} to improve the performance of the vehicular networks \cite{8246040}. A game-theoretic approach was adopted in \cite{xiao2017cooperative} to improve the reliability of message delivery in cooperative vehicular networks by cooperative piggybacking. The simulation results indicated that such an approach help minimize propagation delay while improved broadcast reliability. In \cite{8437267}, Shinde \emph{et al}. use a game-theoretic approach to formulate a Stackelberg game for electric vehicles and utility companies. They found that the lack of competition between utility companies of electric vehicles can lead to monopoly. Thus, to ensure a healthy competition, they employ a distributed algorithm to solve the Stackelberg game resulting in increasing competition between utility companies and lowering the prices. Reference \cite{kadadha2018stackelberg} uses Stackelberg game to minimize the number of hops and maximize the throughput for a multi-hop urban vehicular ad-hoc network. The proposed quality of service (QoS) aware method outperforms the optimized link state routing protocol in terms of throughput and end-to-end delay. However, the performance improvements were discussed only for V2V communication but not V2I communications.

Despite its relentless growth over the last decade, the literature on vehicular communications lacks practical physical layer assumptions. Strictly speaking, it is not uncommon to find the assumption of statistical independence of the radio links at individual antennas of the RSU. As the RSU is generally equipped with closely packed antennas, the assumption of statistical independence of fading links oversimplifies the analysis and cannot provide practical insights. Moreover, to the best of authors' knowledge, results on cooperative communications under correlated fading at the RSU have not been reported yet. Motivated by these observations, our current work makes the following research contributions:
\begin{itemize}
 \item We study the impact of two correlation models for multiple antennas at the RSU, i.e., constant correlation (CC) and exponential correlation (EC). By considering different numbers of antennas at the RSU, we characterize the performance improvements for both the CC model and EC model.
\item We derive closed-form expressions of packet error probability for cooperative vehicular networks in the presence of a multiple-antenna RSU. The links are assumed to be Nakagami-$m$ faded which is a versatile fading model compared to conventionally used Rayleigh fading model \cite{jameel2018impact}.
\item We formulate a game-theoretic model to evaluate optimal transmit power for uplink cooperative vehicular networks. In particular, we consider a non-cooperative Stackelberg game where the source vehicle pays the helper vehicles for forwarding the information to the RSU. Optimal solutions for transmit power and pricing are developed for the proposed game.
\end{itemize}
The rest of the paper is organized as follows. Section II introduces the system model. Section III provides the performance analysis for cooperative vehicular communications. Section IV presents a game-theoretic analysis of the system model. Section V gives numerical results along with their relevant discussion. Finally, Section VI concludes the paper with potential future research directions.
\section{System Model}
\begin{figure}
\centering
\includegraphics[width=.48\textwidth]{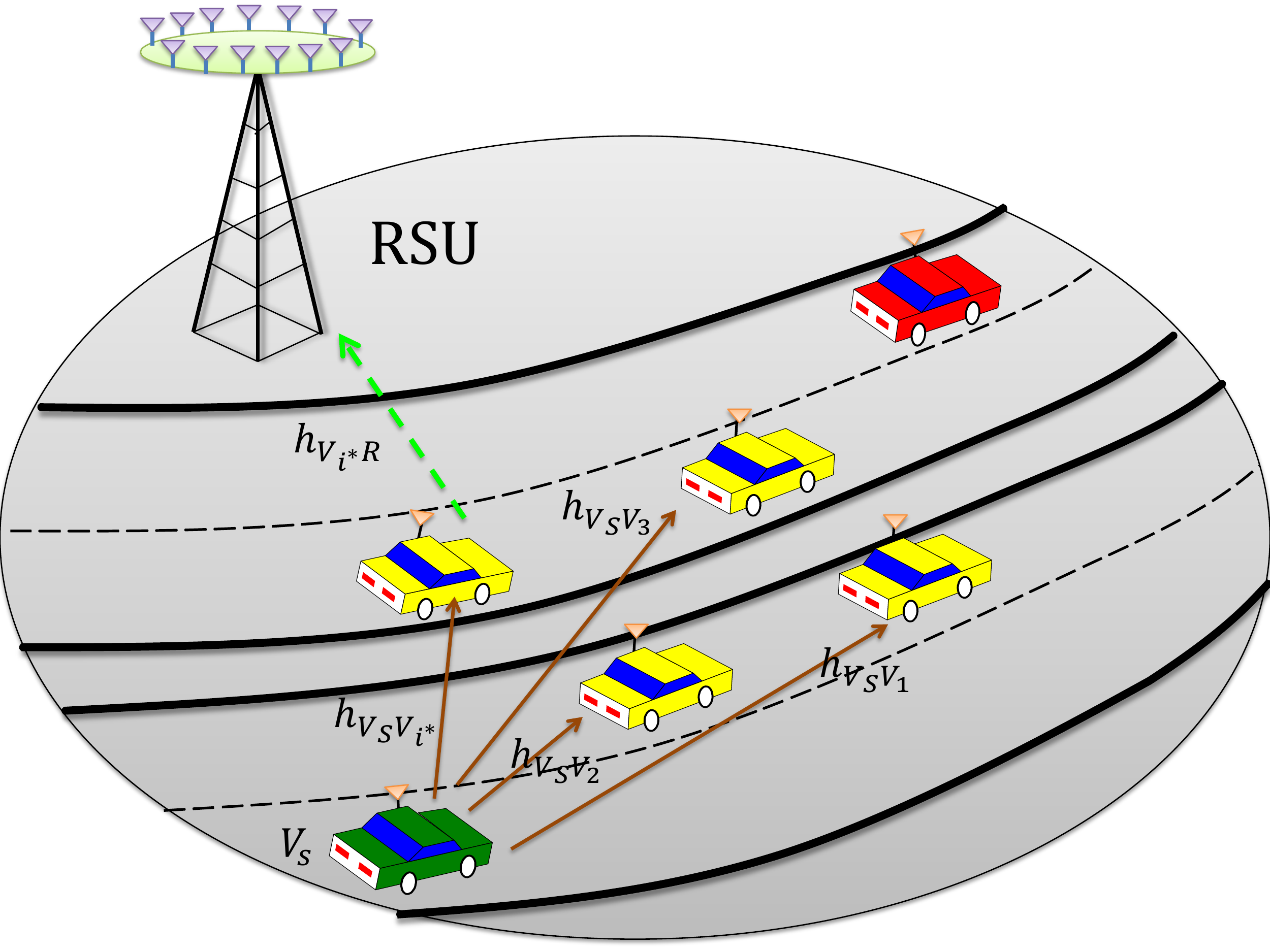}
\caption{System model.}
\label{block}
\end{figure}

In Fig. \ref{block}, we consider a hybrid uplink V2V and V2I system consisting of a source vehicle $V_{s}$, intermediate helper vehicles $\mathcal{V}=\{ V_{i}\vert i=1,2,\ldots N\}$ and an RSU having $M>1$ antennas. Both $V_{s}$ and $V_{i}$ are assumed to be equipped with a single antenna. We assume all links are independent and identically distributed (i.i.d) Nakagami-$m$ faded and follow a block fading model such that the fading during a single block is invariant but changes randomly from one block to another. The transmission takes place in two phases by dividing a single block of time into two-time slots\footnote{For the considered model, the communication is taking place in different time phases and in the presence of a single source vehicle. As such, the co-channel interferences are not incorporated. The analysis for multiple source vehicles is subject of the future work.}. During the first phase, $V_{s}$ broadcasts its signal to a particular $i$-th helper vehicle. The helper vehicle is chosen based on the channel state information (CSI) of the links between $V_{s}$ and intermediate vehicles. We consider that the direct link between the source vehicle and the RSU cannot be used due to high path loss and deep fading. Thus, $V_s$ adopts a more reliable approach for transmitting the message through intermediate helper vehicles. Let $P$ be total transmit power used for communication and $V_{s}$ transmits a signal $x$ to the helper vehicle $V_i$ with transmit power $\varphi P$. In this paper, for simplicity and without loss of generality we separately consider the effects of large-scale pathloss and small-scale fading in our channel model \cite{ploss1, ploss2}. The received signal at $V_{i}$ can then be written as

\begin{align}
y_{V_{s}V_{i}}=\sqrt{\frac{\varphi P}{d_{V_{s}V_{i}}^{\alpha 
}}}h_{V_{s}V_{i}}x+n_{V_{s}V_{i}},
\label{equatee_4}
\end{align}%
where $ h_{V_{s}V_{i}}$ is the fading coefficient between $V_{s}$ and $V_i$, 
$0<\varphi\leq 1$ is the ratio of power used in the first phase, $n_{V_{s}V_{i}}$ is the 
additive white Gaussian noise (AWGN) at $V_{i}$ with zero mean and 
variance $N_{0}$, $d_{V_{s}V_{i}}$ is the distance between $V_{s}
$ and $V_i$, and $\alpha $ is the path loss exponent.   

A helper vehicle is selected by the source vehicle based on the CSI of the first hop. This results in maximizing the signal-to-noise ratio (SNR) between $V_s$ and $V_{i^{*}}$, where ${i^{*}}$ denotes the index of the selected helper vehicle. Using order statistics, it can be written as 
\begin{align}
\gamma _{V_{s}V_{i^{*}}}=\max _{i \in N}\gamma _{V_{s}V_{i}},
\label{equatee_1}
\end{align}
where $\gamma _{V_{s}V_{i}}=\frac{\varphi P}{d_{V_{s}V_{i}}^{\alpha}N_{0}}\vert h_{V_{s}V_{i}}\vert ^{2}$ represents the SNR between the source and helper vehicles.

In the second phase, $V_{i^*}$ decodes the signal and then re-encodes 
it and transmit to the RSU. The signal received at the $j$-th antenna of 
the RSU is given as

\begin{align}
y_{V_{i^{*}}R}^{(j)}=\sqrt{\frac{(1-\varphi) P}{d_{V_{i^{*}}R}^{\alpha 
}}}h_{V_{i^{*}}R}^{(j)}x+n_{V_{i^{*}}R}^{(j)},
\end{align}
where $h_{V_{i^{*}}R}^{(j)}$ is the fading coefficient between $
V_{i^{*}}$ and the RSU, $n_{V_{i^{*}}R}^{(j)}$ is the AWGN at the $j$-th antenna with zero mean and variance $N_{0}$, $d_{V_{i^{*}}R}$ is the distance between $V_{i^{*}}$ and the RSU. Since the RSU is equipped with multiple antennas, a single input multiple output (SIMO) link exists between $V_{i^{*}}$ and RSU. By exploiting this SIMO link, the RSU can employ the maximum ratio combining (MRC) technique to improve its received SNR. The instantaneous SNR at the RSU is then computed as 
\begin{align}
\gamma _{V_{i^{*}}R}=\sum_{j=1}^{M}{\gamma _{V_{i^{*}}R}^{(j)}}, 
\end{align}
where $\gamma _{V_{i^{*}}R}^{(j)}=\frac{(1-\varphi) P}{d_{V_{i^{*}}R}^{\alpha}N_{0}}\vert h_{V_{i^{*}}R}^{(j)}\vert ^{2}$ is the SNR between the helper vehicle and the RSU. It is worthwhile mentioning that the RSU will not be able to receive the message from $V_{i^{*}}$ if the SNR at $V_{i^{*}}$ is not sufficiently high for message decoding. Hence, the end-to-end SNR depends on the bottleneck SNR from $
V_{s}$ to $V_{i^{*}}$ and from $V_{i^{*}}$ to the RSU, which can be 
represented as

\begin{align}
\gamma _{e2e}=\min \lbrace \gamma _{V_{s}V_{i^{*}}},\gamma 
_{V_{i^{*}}R}\rbrace .
\label{eq_e2e}
\end{align}

It should be noted that the amount of energy consumed for receiving and decoding a message is very small compared to the battery capacity of the vehicles. On the other hand, a helper vehicle would in turn benefit from such cooperative communication scheme once it needs other vehicles to assist with its own data transmission.

\section{Packet Error Performance Analysis}

In this section, we derive closed-form expressions of packet error probability for both CC and EC models. The packet error probability is an important metric for reliability analysis of the wireless networks. To counter the effects of error propagation, a vast number of classical error correction and network coding techniques exist in the literature. The packet error probability is also relevant for the analysis of large-scale distributed systems using short data packets. These observations motivate us to derive the closed-form expression of packet error probability for analyzing the performance of vehicular networks. To do so, we derive the packet error probability based on the outage effect of wireless links which can act as a lower bound by assuming the ideal coding \cite{lau2006channel}. As per the previously explained block fading model, the wireless channel remains unchanged during the coherence time. Thus, we divide a packet into $L$ blocks, wherein the number of blocks depends on the vehicle speed. Here, $L$ is expressed as 
\begin{align}
L=\frac{\Psi }{T_{c}\log (1+\gamma _{0})},
\label{eq:6}
\end{align}%
where $\Psi $ is the size of packet, $T_{c}=\frac{3cf_{c}}{4\sqrt{\pi }(c+v)}$ is the coherence time, $c$ is the speed of light, $f_{c}$ is the carrier frequency, $v$ is the vehicle speed and $\gamma_{0}$ is the SNR threshold for successful decoding. Note that the coherence time is inversely proportional to the Doppler spread. As the vehicle speed increases, the Doppler spread also increases resulting in smaller coherence time. This observation is consistent with the above expression of $T_c$. Moreover, although the 802.11 systems are able to adaptively select the data rates based on the distance between source and destination, we use a constant value of $\gamma _{0}$ for the sake of mathematical tractability and without loss of generalization. By using order statistics and exploiting the independence of $\gamma_{V_{s}V_{i^{*}}}$ and $\gamma _{V_{i^{*}}R}$ the packet error probability for the $l$-th block can be written as

\begin{align}
P_{err,l}&=\Pr(\gamma _{e2e}<\gamma _{0}) =\Pr(\min \lbrace \gamma _{V_{s}V_{i^{*}}},\gamma 
_{V_{i^{*}}R}\rbrace <\gamma _{0}) \nonumber \\
&=\Pr(\gamma _{V_{s}V_{i^{*}}}<\gamma _{0})+\Pr(\gamma 
_{V_{i^{*}}R}<\gamma _{0})\nonumber \\
&-\Pr(\gamma _{V_{s}V_{i^{*}}}<\gamma 
_{0}) \times \Pr(\gamma _{V_{i^{*}}R}<\gamma _{0}).
\label{eq_8}
\end{align}

The probability of $\gamma _{V_{s}V_{i^{*}}}$
falling below $\gamma _{0}$ using (\ref{equatee_1}) can be simplified as 
\begin{align}
\Pr(\gamma _{V_{s}V_{i^{*}}}<\gamma _{0})=\Pr(\max _{i \in N}\gamma _{V_{s}V_{i}}<\gamma_0)
\label{equatee_2}
\end{align}

Due to large separation of helper vehicles, the channels between source and helper vehicles are considered to experience independent fading. Thus, (\ref{equatee_2}) can be re-written as
\begin{align}
\Pr(\gamma _{V_{s}V_{i^{*}}}<\gamma _{0})&=\Pr(\gamma _{V_{s}V_{1}}<\gamma_0)\times\Pr(\gamma _{V_{s}V_{2}}<\gamma_0) \nonumber \\
&\times\Pr(\gamma _{V_{s}V_{3}}<\gamma_0)...\times \Pr(\gamma _{V_{s}V_{N}}<\gamma_0)\nonumber \\
&= \prod_{i=1}^{N}{\Pr(\gamma_{V_{s}V_{i}}<\gamma _{0})}.
\label{equatee_3}
\end{align}

As the links are Nakagami-$m$ distributed, the SNR will be Gamma distributed with the probability density function (PDF) given as \cite{he2016performance}

\begin{align}
f_{Z}(z)=\left(\frac{m}{\bar{z}}\right)^{m}\frac{z^{m-1}}{\Gamma 
(m)}\exp\left(-\frac{ms}{\bar{z}}\right), 
\label{pdf}
\end{align}
where $m\ge \frac{1}{2}$ is the Nakagami-$m$ parameter. Here, $m=1$ represents 
Rayleigh fading and $m=\infty $ corresponds to a nonfading channel. Also, $\bar{z}$ is the mean of the distribution and $\Gamma (.)$ denotes the Gamma function. By using (\ref{pdf}) and simplifying the integral, we arrive at 

\begin{align}
&\Pr(\gamma _{V_{s}V_{i^{*}}}<\gamma 
_{0})=\prod_{i=1}^{N}{\frac{\gamma \left(m,\frac{\gamma _{0}}{\bar{\gamma }_{V_{s}V_{i}}}\right)}{\Gamma (m)}},
\label{eq_12}
\end{align}
where $\gamma(.,.)$ is the lower incomplete Gamma function and $\bar{\gamma }_{V_{s}V_{i}}$ is the mean SNR. Next, we calculate the probability of $\gamma _{V_{i^{*}}R}$ falling below the $\gamma _{0}$, which is given as

\begin{align}
\Pr(\gamma _{V_{i^{*}}R}<\gamma _{0})=\Pr\left(\sum_{j=1}^{M}{\gamma 
_{V_{i^{*}}R}^{(j)}}<\gamma _{0}\right).
\label{eq_1}
\end{align}

We consider that the antennas are crowded at the RSU and hence experience correlated fading due to the minimal antenna separation. To incorporate the effect of correlation in our considered system, we will analyze two antenna correlation 
models, namely, CC and EC.

Let us first consider the case of CC where the value of the correlation coefficient $\rho _{c}$ remains unchanged despite a change in the distance of the closely packed antennas. In this case, the PDF of the received 
SNR at the output of the combiner under Nakagami-$m$ fading is given by \cite{gurland1955distribution}

\begin{align}
f_{Z}(z)=&\left(\frac{zm}{\bar{z}}\right)^{Mm-1}\exp\left(-\frac{zm}{\bar{z}(1-\rho 
_{c})}\right) \nonumber \\
&\times \frac{_{1}F_{1}\left(m,Mm;\frac{Mm\rho _{c}z}{\bar{z}(1-\rho _{c})(1-\rho 
_{c}+M\rho _{c})}\right)}{(\frac{\bar{z}}{m})(1-\rho _{c})^{m(M-1)}(1-\rho 
_{c}+M\rho _{c})^{m}\Gamma (Mm)},
\label{pdf2}
\end{align}
where $_1F_1(.)$ is the confluent hypergeometric function. By substituting (\ref{pdf2}) in (\ref{eq_1}) and with the help of \cite[Eq. (9.111)]{gradshteyn2014table}, we obtain

\begin{align}
\Pr(\gamma _{V_{i^{*}}R}<\gamma _{0})&=\frac{1}{\Gamma (m)\Gamma 
(v)}\left(\frac{1-\rho _{c}}{M\rho _{c}} \right)^{m}\left(\frac{1-\rho _{c}+M\rho 
_{c}}{M\rho _{c}}\right)^{v} \nonumber \\
&\times \int_{0}^{\frac{M\rho _{c}m\gamma 
_{0}}{\bar{\gamma }_{V_{i^{*}}R}(1-\rho _{c})(1-\rho _{c}+M\rho 
_{c})}}{\int_{0}^{1}{\gamma _{V_{i^{*}}R}^{Mm-1}t^{m-1}}} \nonumber \\
&\times (1-t)^{v-1} \times \exp\biggl\{-\left(\frac{1-\rho _{c}+M\rho _{c}}{M\rho _{c}}-t\right) \nonumber \\
& \times \gamma_{V_{i^{*}}R}\biggr\}\times dt d\gamma _{V_{i^{*}}R},
\label{eqeq}
\end{align}
where $v=Mm-m$. By using the identities \cite[Eqs. (3.385) \& (9.261)]{gradshteyn2014table}, the integrals in (\ref{eqeq}) can be resolved as 

\begin{align}
\Pr(\gamma _{V_{i^{*}}R}<\gamma _{0})&=\frac{1}{\Gamma (m)\Gamma 
(v)}\left(\frac{1-\rho _{c}}{M\rho _{c}} \right)^{m}\left(\frac{1-\rho _{c}+M\rho 
_{c}}{M\rho _{c}}\right)^{v} \nonumber \\
&\times \Phi _{1}\left(m,Mm,Mm,\frac{M\rho _{c}}{1-\rho 
_{c}+M\rho _{c}},0\right) \nonumber \\
&-\frac{1}{\Gamma (m)\Gamma (v)} \left(\frac{1-\rho _{c}}{M\rho _{c}} 
\right)^{m} \left(\frac{1-\rho _{c}+M\rho _{c}}{M\rho _{c}}\right)^{v}\nonumber \\
&\times \exp\left(-\frac{(1-\rho 
_{c}+M\rho _{c})\gamma _{0}}{M\rho _{c}}\right) \hspace{-0.5cm}\sum_{n=0}^{\frac{1-\rho _{c}+M\rho _{c}}{M\rho _{c}}-1} {\frac{(\gamma _{0})^{n}}{n!}} \nonumber \\
&\times \Phi _{1}\biggl(m,Mm-n,Mm,\frac{M\rho _{c}}{1-\rho _{c}+M\rho 
_{c}}, \nonumber \\
&\frac{M\rho _{c}m\gamma _{0}}{\bar{\gamma }_{V_{i^{*}}R}(1-\rho 
_{c})(1-\rho _{c}+M\rho _{c})}\biggr),
\label{eq_cc}
\end{align}
where $\Phi _{1}(.)$ is the generalized hypergeometric function and $\bar{\gamma }_{V_{i^{*}}R}$ is the mean SNR from the selected helper vehicle to the RSU. The packet error probability for $l$-th block when the RSU assumes the CC model can be obtained by substituting (\ref{eq_cc}) and (\ref{eq_12}) into (\ref{eq_8}).

For the case of EC, we consider that the correlation between the signals increases with the decrease in spatial separation between two antennas. The PDF of the received SNR at the output of the combiner for Nakagami-$m$ faded links becomes \cite{kotz1964distribution} 

\begin{align}
f_{Z}(z)=\frac{z^{\frac{mM^{2}}{\lambda }-1}\exp(-\frac{Mmz}{\lambda 
\bar{z}})}{\Gamma (\frac{mM^{2}}{\lambda })(\frac{\lambda 
\bar{z}}{Mm})^{\frac{mM^{2}}{\lambda }}},
\label{pdf3}
\end{align}
where $\lambda =M+\frac{2\rho _{e}}{1-\rho _{e}}(M-\frac{1-\rho 
_{e}^{M}}{1-\rho _{e}})$ and $\rho _{e}$ is the correlation 
coefficient for the EC model. By using (\ref{pdf3}) along with 
(\ref{eq_1}) and after a variable transformation, we obtain 

\begin{align}
\Pr(\gamma 
_{V_{i^{*}}R}<\gamma _{0})=\int_{0}^{\gamma _{0}}{\frac{\gamma 
_{V_{i^{*}}R}^{\frac{mM^{2}}{\lambda }-1}\exp(-\frac{Mm\gamma 
_{V_{i^{*}}R}}{\lambda \bar{\gamma }_{V_{i^{*}}R}})}{\Gamma 
(\frac{mM^{2}}{\lambda })(\frac{\lambda \bar{\gamma 
}_{V_{i^{*}}R}}{Mm})^{\frac{mM^{2}}{\lambda }}}}d\gamma _{V_{i^{*}}R}.
\end{align}

With the help of \cite[Eq. (8.350)]{gradshteyn2014table} and after some algebraic simplifications, we get
\begin{align}
\Pr(\gamma _{V_{i^{*}}R}<\gamma _{0})=\Gamma \left(\frac{mM^{2}}{\lambda 
},\frac{Mm\gamma _{0}}{\lambda \bar{\gamma }_{V_{i^{*}}R}}\right),
\label{eq_ec}
\end{align}
where $\Gamma(.,.)$ is the upper incomplete Gamma function. We can get the packet error probability for the $l$-th block in the EC model by a straightforward substitutions of (\ref{eq_12}) and (\ref{eq_ec}) into (\ref{eq_8}). Finally, the packet error probability for all $L$ blocks can be obtained as

\begin{align}
P_{err}=1-(1-P_{err,l})^{L},
\end{align}
where $P_{err,l}$ is obtained from (\ref{eq_8}).

\section{Game-Theoretic Analysis}

In this section, we formulate a game theoretic model with the goal to derive an optimal transmit power strategy for the considered cooperative network. Following the system model in Section II, the transmission takes place in two phases. In the first phase, the source vehicle selects a helper vehicle among a set of vehicles and transmits the message to the selected helper vehicle. In the second phase, the selected vehicle decodes the received message and transmits the re-encoded message to the RSU. The information asymmetry between the source and helper vehicles, the non-cooperative selection of the helper vehicle, and the sequential nature of the end-to-end communication motivate us to apply the Stackelberg game on our system model.

Stackelberg game is a sequential non-cooperative game, where players are required to make decision hierarchically. The players are divided into two sets of players, i.e., leaders and followers \cite{kadadha2018stackelberg}. The leaders hold a strong position based on some pre-specified criteria, whereas the rest of the players are followers. This dominant position of leaders also leads to an asymmetry of information among leaders and followers. The leaders declare their strategy first and, due to hierarchical decision-making, they can enforce their strategies on the followers \cite{han2012game}. The followers react to the strategies of leaders, wherein they may play a non-cooperative game among themselves. As a special case, the Stackelberg game can be easily extended for the single leader and multiple followers scenario. In this case, the leader only defines a single reaction while the followers maximize their utilities by calculating the optimal response.

While analyzing the model in Section II, we observe that the communication patterns between the source and helper vehicles follow a leader-followers model. Since the source vehicle initiates the communication and, subsequently, selects one of the helper vehicle, the strategy of helper vehicles is dependent on the decision of the source vehicles. This shows the dominance of the source vehicle and favors it to become the leader in this game. Due to the influence of source vehicle, the optimal strategy of helper vehicles would be determined by the initial response of the source vehicle, making helper vehicles the followers in this game. The source vehicle, being the leader in this game, is considered to have the advantage of selecting the values of $\varphi$ and $P$ to maximize its own utility. Based on these values, the helper vehicles play a non-cooperative game among themselves. In this way, each helper vehicle reacts to the already decided values of $\varphi$ and $P$ by deciding the payment it is willing to accept. It is observed that the communication scenario is similar to a leader-follower 
game and thus can be analyzed using the Stackelberg game.

First, we will define the utility function of the source vehicle which determines the degree of satisfaction of the vehicle. The main objective of the source vehicle is to ensure the reliability of sent messages. In other words, the source vehicle is interested in the SNR of the received signal at the RSU, thus, the satisfaction of the source vehicle can be considered as a sigmoid function of the end-to-end SNR
\begin{align}
U_{R}=\frac{1}{1+\exp\left\{-a(\gamma _{e2e}-\gamma _{0})\right\}},
\label{equatee_5}
\end{align}
where $a$ denotes the steepness of the satisfaction curve and $\gamma_{0}$ is the SNR threshold requirement of the source vehicle. It is worth mentioning that the sigmoid function has been extensively used to model user's satisfaction with respect to resource allocation and service qualities \cite{lin2005arc,stamoulis1999efficient}. The value the SNR threshold is an indicator of the error rate of the received message. In other words, if the received SNR is below the value of $\gamma _{0}$, the source vehicle has poor satisfaction level, whereas, the satisfaction rapidly increases when SNR is significantly higher than $\gamma _{0}$. Aside from the SNR threshold, the utility function of the source vehicle is also dependent on the price set by the helper vehicles. It is obvious that the source vehicle may not be willing to pay any price just to ensure the reception of the message at RSU. As a result of this, the net utility function of the source vehicle is a weighted sum of the utility function of SNR satisfaction and the revenue that the helper vehicle collects from the source vehicle. It can be represented as 
\begin{align}
U_{s}=w_{p}U_{R}-H_{i},
\label{eq_us}
\end{align}
where $w_{p}$ is a predefined parameter in the unit of revenue 
per SNR utility, $H_{i}=p_{i}(1-\varphi )P$ is the cost paid by 
the source vehicle to the selected helper vehicle and $p_{i}$ is 
the price unit per each Watt of power set by the selected helper vehicle. 

It is worth pointing out that the helper vehicles also target maximizing their profit under a reasonable cost. More specifically, each helper vehicle tries to earn a payment from the source vehicle to gain most of the profit while covering the forwarding cost of the signal. The utility function of the $i$-th helper vehicle is given as 

\begin{align}
U_{H}=(p_{i}(1-\varphi )-c_{i})P,
\label{eq_uh}
\end{align}
where $c_{i}$ is the cost per unit power incurred by the helper 
the vehicle in forwarding the signal to the RSU. Without loss of generality, we 
assume $c_{i}=c ~ \forall \ i \ \in \mathcal{V}$. Eqs (\ref{eq_us}) and (\ref{eq_uh}) show that if a helper vehicle asks a higher price then the source vehicle can buy less from that 
helper vehicle and it can even completely disregard the services of that helper. In contrast, 
if the price is too low, then the profit received by helper vehicle would be unnecessarily low, which may not be acceptable to the helper vehicle.

\begin{prop}
$U_{s}$ is maximized if and only if $\varphi ^{*}=\frac{\eta d_{V_{s}V_{i}}^{\alpha }}{\eta 
d_{V_{s}V_{i}}^{\alpha }+\vert h_{V_{s}V_{i}}\vert 
^{2}d_{V_{i^{*}}R}^{\alpha }}$.
\end{prop}

\begin{IEEEproof}
From (\ref{eq_us}), maximization of $U_{s}$ is 
based on the maximization of $U_{R}$ which increases as the value of $\gamma _{e2e}$ increases in sigmoid function in (\ref{equatee_5}). From (\ref{eq_e2e}), it can be seen that $\gamma _{e2e}$ is minimum of the increasing function of $\gamma _{V_{s}V_{i^{*}}}$ and a decreasing function of $\gamma_{V_{i^{*}}R}$. Thus, the maximization of $U_s$ is achieved when $\gamma _{V_{s}V_{i^{*}}}=\gamma_{V_{i^{*}}R}$. Solving for $\varphi$ yields

\begin{align}
\varphi ^{*}=\frac{\eta d_{V_{s}V_{i}}^{\alpha }}{\eta 
d_{V_{s}V_{i}}^{\alpha }+\vert h_{V_{s}V_{i}}\vert 
^{2}d_{V_{i^{*}}R}^{\alpha }},
\end{align}
where $\eta =\sum_{j=1}^{M}{\vert h_{V_{i^{*}}R}^{(j)}\vert ^{2}}$.
\end{IEEEproof}

Now (\ref{eq_us}) can be re-written as

\begin{align}
U_{s}=&\frac{w_{p}}{1+e^{-a\left(\frac{\eta d_{V_{s}V_{i}}^{\alpha }}{\eta 
d_{V_{s}V_{i}}^{\alpha }+\vert h_{V_{s}V_{i}}\vert 
^{2}d_{V_{i^{*}}R}^{\alpha }}\times \frac{P\vert h_{V_{s}V_{i}}\vert 
^{2}}{d_{V_{s}V_{i}}^{\alpha }N_{0}}-\gamma _{0}\right)}} \nonumber \\
&-p_{i}\left(1-\frac{\eta d_{V_{s}V_{i}}^{\alpha }}{\eta d_{V_{s}V_{i}}^{\alpha }+\vert 
h_{V_{s}V_{i}}\vert ^{2}d_{V_{i^{*}}R}^{\alpha }}\right)P.
\label{eq_us2}
\end{align}

\begin{prop}
If the selling price $p_{i}$ is given, then $\frac{\partial U_{s}}{\partial P}=0$ is at optimality.
\end{prop}

\begin{IEEEproof}
It can be observed from (\ref{eq_us2}) that when $P$ is close to 0, $U_{s}$ is close to 
0 and little help is received from the helper vehicle. With the increase in the value of $P$, the helper vehicle sells more power to the source vehicle so a large increment is obtained in the received SNR. As the value of $P$ increases further, the cost of transmission will grow but the received SNR 
will saturate and the utility of $U_{s}$ will begin to decrease. Thus,
by calculating the first order derivative with respect to $P$, we 
have 

\begin{align}
\frac{\partial U_{s}}{\partial P}=&\frac{a\eta \vert 
h_{V_{s}V_{i}}\vert ^{2}w_{p}e^{-a\left(-\gamma _{0}+\frac{\eta P\vert 
h_{V_{s}V_{i}}\vert ^{2}}{\varpi}\right)}}{\varpi\left\{ 1+e^{-a\left(-\gamma _{0}+\frac{\eta 
P\vert h_{V_{s}V_{i}}\vert ^{2}}{\varpi}\right)}\right\} 
}-p_{i} \nonumber \\
&\times \left(1-\frac{\eta d_{V_{s}V_{i}}^{\alpha }}{\eta 
d_{V_{s}V_{i}}^{\alpha }+\vert h_{V_{s}V_{i}}\vert 
^{2}d_{V_{i^{*}}R}^{\alpha }}\right),
\label{eq_us3}
\end{align}
where $\varpi=N_{0}\left(\eta d_{V_{s}V_{i}}^{\alpha }+\vert 
h_{V_{s}V_{i}}\vert ^{2}d_{V_{i^{*}}R}^{\alpha }\right)$. Now, taking a further derivative of 
(\ref{eq_us3}) yields

\begin{align}
\frac{\partial^2 U_{s}}{\partial P^2}=&-\frac{a^{2}\eta ^{2}e^{-a\left(-\gamma _{0}+\frac{\eta \vert 
h_{V_{s}V_{i}}\vert ^{2}P}{\varpi}\right)}\vert 
h_{V_{s}V_{i}}\vert ^{4}w_{p}}{\left\{1+e^{-a\left(-\gamma _{0}+\frac{\eta \vert 
h_{V_{s}V_{i}}\vert ^{2}P}{\varpi}\right)}\right\}^{2}\varpi^2} \nonumber \\
&+\frac{2a^{2}\eta ^{2}e^{-2a\left(-\gamma 
_{0}+\frac{\eta \vert h_{V_{s}V_{i}}\vert ^{2}P}{\varpi}\right)}\vert h_{V_{s}V_{i}}\vert 
^{4}w_{p}}{\left \{1+e^{-a\left(-\gamma _{0}+\frac{\eta \vert h_{V_{s}V_{i}}\vert 
^{2}P}{\varpi}\right)}\right\}^{3}\varpi^{2}}.
\end{align}

Note that $\frac{\partial^2 U_{s}}{\partial P^2}$ is always less than 0 for $P$,$\gamma_0>0$ and $0<h_{V_{s}V_{i}},w_p<1$. Therefore, $U_{s}$ is concave in $P$ and the optimal power can be obtaining by solving $\frac{\partial U_{s}}{\partial P}=0$.
\end{IEEEproof}

Solving $\frac{\partial U_{s}}{\partial P}$ for $P$ gives
\begin{align}
\bar{P}=&\frac{\varpi \log \left(\frac{-2N_{0}p_{i}d_{V_{i^{*}}R}^{\alpha }+a\eta 
w_{p}+\sqrt[]{-4a\eta N_{0}p_{i}d_{V_{i^{*}}R}^{\alpha }w_{p}+(a\eta 
w_{p})^{2}}}{2N_{0}p_{i}d_{V_{i^{*}}R}^{\alpha }}\right) }{a\eta \vert 
h_{V_{s}V_{i}}\vert ^{2}} \nonumber \\
&+\frac{\varpi \gamma _{0}}{\eta \vert h_{V_{s}V_{i}}\vert ^{2}}.
\end{align}

\begin{figure}[!htp]
\centering
\includegraphics[width=.53\textwidth]{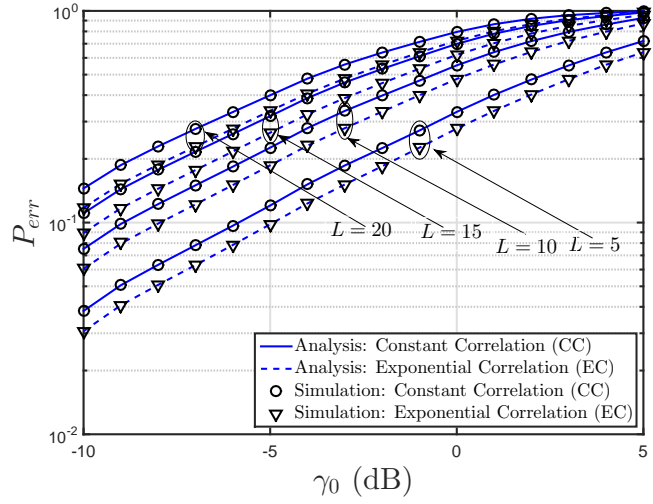}
\caption{$P_{err}$ as a function of $\gamma_0$, for CC and EC models.}
\label{fig22}
\end{figure}

\begin{figure}[!htp]
\centering
\includegraphics[width=.53\textwidth]{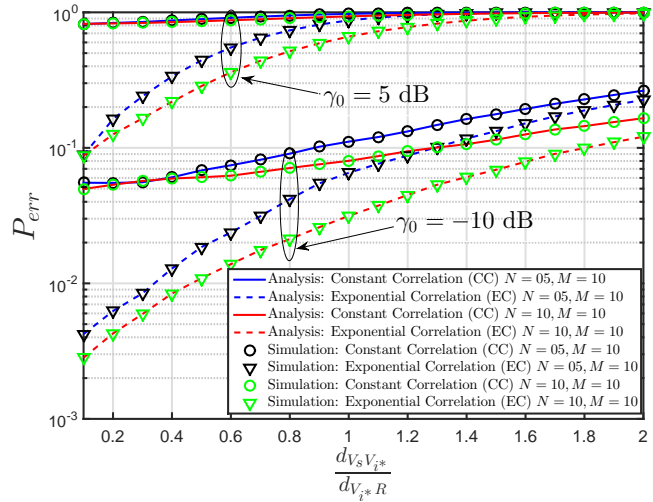}
\caption{$P_{err}$ versus $\frac{d_{V_sV_{i^*}}}{d_{V_{i^*}R}}$ for different values of $N$ and $M$.}
\label{fig33}
\end{figure}

As the above solution can be negative for some values of $p_{i}$, the optimal price is set 
as $P^{*}=\max (\bar{P},0)$. In order to obtain the optimal price value for the $i$-th helper vehicle, we first differentiate (\ref{eq_uh}) with respect to $p_i$, as given in (\ref{eq_top}) where the approximation comes from the Taylor series expansion. The optimal price for the helper vehicle can be obtained by solving $\frac{\partial U_{H}}{\partial p_{i}}=0$ as
\begin{figure*}
\centering
\begin{align}
\frac{\partial U_{H}}{\partial p_{i}} \approx & 
\frac{2N_{0}p_{i}d_{V_{i^{*}}R}^{\alpha } \varpi \left\{-c+p_{i}(1-\frac{\eta d_{V_{s}V_{i}}^{\alpha }}{\eta 
d_{V_{s}V_{i}}^{\alpha }+\vert h_{V_{s}V_{i}}\vert 
^{2}d_{V_{i^{*}}R}^{\alpha }})\right\}}{\left\{a\eta \vert h_{V_{s}V_{i}}\vert 
^{2}(-2N_{0}p_{i}d_{V_{i^{*}}R}^{\alpha }+a\eta w_{p}+\sqrt{-4a\eta 
N_{0}p_{i}d_{V_{i^{*}}R}^{\alpha }w_{p}+a^{2}\eta ^{2}w_{p}^{2}})\right\}}\biggl(\frac{-2N_{0}d_{V_{i^{*}}R}^{\alpha }-\frac{2a\eta 
N_{0}d_{V_{i^{*}}R}^{\alpha }w_{p}}{\sqrt{-4a\eta 
N_{0}p_{i}d_{V_{i^{*}}R}^{\alpha }w_{p}+a^{2}\eta 
^{2}w_{p}^{2}}}}{2N_{0}p_{i}d_{V_{i^{*}}R}^{\alpha 
}}\nonumber \\
&-\frac{-2N_{0}p_{i}d_{V_{i^{*}}R}^{\alpha }+a\eta w_{p}+\sqrt{-4a\eta 
N_{0}p_{i}d_{V_{i^{*}}R}^{\alpha }w_{p}+a^{2}\eta 
^{2}w_{p}^{2}}}{2N_{0}p_{i}^{2}d_{V_{i^{*}}R}^{\alpha }}\biggr).
\label{eq_top}
\end{align}
\hrulefill
\end{figure*}
\begin{align}
p_{i}^{*}= \frac{\eta cd_{V_{s}V_{i}}^{\alpha }+c\vert 
h_{V_{s}V_{i}}\vert ^{2}d_{V_{i^{*}}R}^{\alpha }}{\vert 
h_{V_{s}V_{i}}\vert ^{2}d_{V_{i^{*}}R}^{\alpha }}.
\end{align}
\section{Numerical Results}

This section provides analytical and simulation results based on the mathematical analysis in Sections III \& IV. We perform link level simulations in MATLAB for $10^5$ channel realizations. Unless stated otherwise, the following values are used: $\gamma_0=-10$ dB, $N=5$, $M=10$, $\varphi=0.5$, $m=1$, SNR=$\frac{P}{N_0}=25$ dB, $\rho_c=\rho_e=0.1$, $L=10$. 

In Fig. \ref{fig22}, we plot the packet error probability for different values of SNR threshold. The error probability of received packets increases with the increase in $\gamma_0$. Also, for $\gamma_0=-5$ dB, the packet error probability for the EC model increases from 0.09 to 0.4 as $L$ increases from 5 to 20. This shows that packet error probability increases if the coherence time is small, i.e., the packet is divided into multiple blocks. It can also be seen that the curves of different values of $L$ converge for higher values of $\gamma_0$. This result implies that the impact of coherence time on packet error probability is reduced when the SNR threshold is high, for both EC and CC models.  

Fig. \ref{fig33} illustrates the change in the packet error probability for increasing values of the distance ratio $\frac{d_{V_sV_{i^*}}}{d_{V_{i^*}R}}$. It is clear that an increase in $\frac{d_{V_sV_{i^*}}}{d_{V_{i^*}R}}$ causes an increase in the packet error probability, which can be attributed to the large decoding errors at $V_{i^*}$. Moreover, at lower values of $\frac{d_{V_sV_{i^*}}}{d_{V_{i^*}R}}$, the EC model shows a significant reduction in the packet error probability, while both EC and CC curves saturate for the higher values of distance ratio. The convergence of EC and CC curves is more prominent for larger values of $\gamma_0$. This result indicates the reduction in the impact of the number of helper vehicles at the packet error probability. Also note that for the same values of $M$, an increase in the number of helper vehicles results in decreasing the $P_{err}$. This observation indicates the improved diversity gains obtained by introducing more helper vehicles in the network.

\begin{figure}[!htp]
\centering
\includegraphics[width=.52\textwidth]{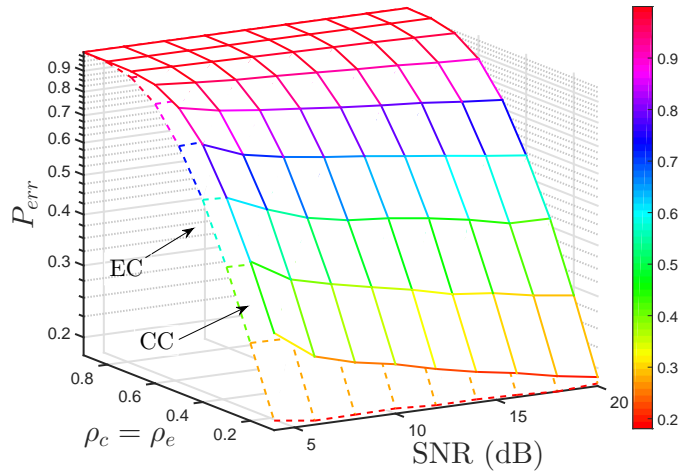}
\caption{$P_{err}$ against SNR and $\rho_e=\rho_c$.}
\label{fig44}
\end{figure}

Fig. \ref{fig44} emphasizes the significance of antenna correlation by plotting $P_{err}$ for different values of SNR and $\rho_e=\rho_c$. The obtained result conforms with our previous results where $P_{err}$ drops with an increase in SNR. In addition to this, we observe that the packet error probability increases with the increase in correlation coefficients $\rho_c$ and $\rho_e$. This trend indicates that a higher antenna correlation causes a large number of packet errors. Moreover, we note that both the EC and CC models converge as $\rho_c$ and $\rho_e$ approach 1. The joint effect of SNR and $\rho_c=\rho_e$ on both EC and CC models can also be seen from the plots. In particular, for 20 dB SNR, the difference between packet error probabilities of both EC and CC reduces as $\rho_c=\rho_c \to 0$. This observation confirms that SNR has a prominent impact on packet error probability for lower values of antenna correlation coefficients.    
\begin{figure}[!htp]
\centering
\includegraphics[width=.53\textwidth]{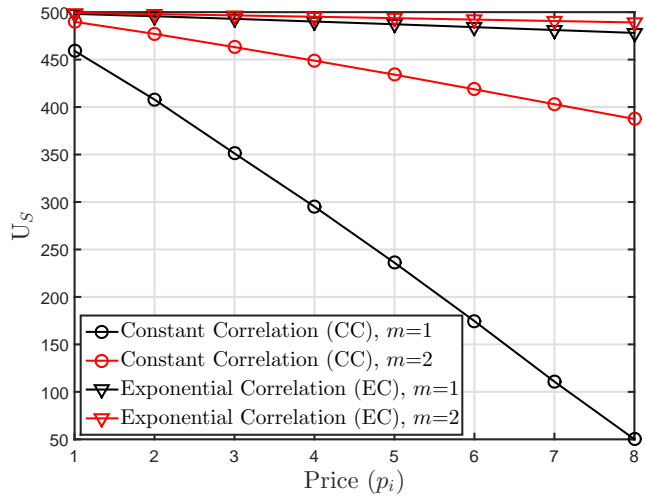}
\caption{Source vehicle utility function versus increasing price value.}
\label{fig55}
\end{figure}

Fig. \ref{fig55} illustrates the impact of different values of price on the utility function of the source vehicle. It can be seen that for a small price asked by the helper vehicle, the source vehicle is inclined to get more power in order to improve its SNR of the received message. However, as the asking price increases beyond the payment ability of the source vehicle, the source vehicle buys less power which in turn reduces its utility. Additionally, we note that the wireless channel has a noteworthy impact on the utility function of the source vehicle. As the severity of fading increases, i.e., the value of $m$ reduces from 2 to 1, the utility function rapidly decreases, especially for the CC model. 

\begin{figure}[!htp]
\centering
\includegraphics[width=.52\textwidth]{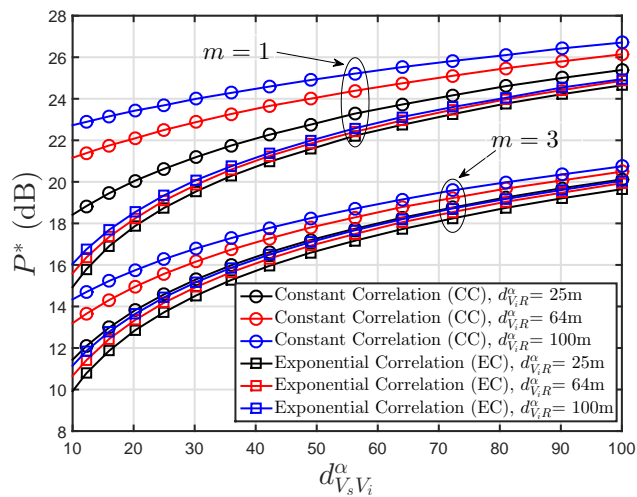}
\caption{$P^*$ as a function of $d_{V_sV_i}$, where $p_i=10$ and $M=5$.}
\label{fig77}
\end{figure}

Fig. \ref{fig77} shows that the optimal value of transmit power increases with the increase in $d_{V_sV_i}$ for both CC and EC models. Moreover, it also increases when the helper vehicle and the RSU are farther from each other. For the CC model, when $d_{V_sV_i}$=10m, the optimal power increases from 18 dB to 23 dB as $d_{V_sR}$ increases from 25m to 100m. Nevertheless, this increase is less prominent for the EC model which indicates that it is least affected by the distance between transmitter and receiver. In any case, for higher values of $d_{V_sV_i}$ all the curves converge, illustrating the diminishing effect of $d_{V_sR}$. This is because the ability of the helper vehicle to decode the message is hampered when $d_{V_sV_i}$ is significantly large, resulting in the requirement of large transmit power to satisfy the SNR threshold $\gamma_0$ at the receiver. Additionally, with an improvement in channel conditions (i.e., an increase in Nakagami-$m$ factor), the value of $P^*$ decreases for both CC and EC models.

\section{Conclusions}%
In this paper, we have provided a realistic evaluation of packet error probability by considering the effect of antenna correlation at an RSU. We have presented the uplink analytical model where intermediate helper vehicles assist in forward dissemination of the source vehicle message to the RSU. We have then derived closed-form analytical expressions of packet error probability under Nakagami-$m$ fading and illustrated the impact of fading parameter $m$ on the packet error probability. Our results show that packet error probability of the EC model resolves to the CC model at higher values of $m$ and correlation coefficients. We have also performed practical analysis by formulating a Stackelberg game where the source vehicle has to pay the helper vehicle for message forwarding to RSU. The obtained numerical results have shown that when the asking price by the helper vehicle is low, the source vehicle is inclined to buy more power to improve its utility function. We have also noted that the optimal power value is dependent on the channel state and the distance between vehicles for both CC and EC models. Our results can be of significant importance for realistic performance evaluation of uplink cooperative vehicular networks.

In this paper, we have considered the case where perfect knowledge of channel state is available to select a helper vehicle. However, due to feedback delays and hardware limitations, it may not always be possible to perfectly estimate the channel conditions. These imperfections may have a degrading effect on the communications system. In the future, we aim to quantify the impact of imperfect CSI on the performance of cooperative vehicular networks. 
\section*{Acknowledgment}%
The authors would like to acknowledge the valuable comments and encouragement received from Prof. Sabita Maharjan. 

\bibliographystyle{IEEEtran}
\bibliography{mybibfile}

%




\end{document}